\documentclass[11pt]{article}
\usepackage{geometry}                
\usepackage[parfill]{parskip}    
\usepackage{graphicx}
\usepackage{amsmath, amsthm, amsfonts}
\usepackage{amssymb}
\usepackage{cite}

\usepackage{color}
     \usepackage[colorlinks]{hyperref}
\hypersetup{linkcolor=blue,%
citecolor=blue,%
urlcolor=cyan}
\usepackage{url} 

\usepackage{epstopdf}
\title{\bf The role of the effective mass in\\ 
two-dimensional Dirac 
electric quantum dots}

\author{
\c{S}. Kuru$^1$\footnote{sengul.kuru@science.ankara.edu.tr, ORCID: \href{http://orcid.org/0000-0001-6380-280X}{0000-0001-6380-280X}}, 
J. Negro$^2$\footnote {jnegro@fta.uva.es, ORCID: \href{http://orcid.org/0000-0002-0847-6420}{0000-0002-0847-6420}}, 
and S. Salamanca$^2$\footnote {sergio.salamanca@uva.es, ORCID: \href{http://orcid.org/0000-0003-0151-8373}{0000-0003-0151-8373}}
\medskip
\\ 
\small
\noindent
$^1$\,Department of Physics, Faculty of Science, Ankara
University, 06100 Ankara, Turkey
\\
\small
\noindent
$^2$\,Departamento de F\'{\i}sica Te\'orica, At\'omica y
\'Optica and Laboratory for Disruptive
\\
\small
\noindent
Interdisciplinary Science (LaDIS), Universidad de Valladolid,  47011 Valladolid, Spain
}

\begin{document}

\maketitle

\begin{abstract}
%

We investigate the influence of a different effective mass inside and outside an electric quantum dot on in its energy spectrum. Depending on the different values we give to the mass, we have found quite different spectra. Specifically, when the mass is positive but lighter inside the dot than outside it, the spectrum increases and splits into two types of states separated by a gap. Conversely, if the mass inside the quantum dot is heavier than the mass outside, the spectrum has fewer states and needs stronger fields in order to confine states. Finally, the case of inverted mass (ie, negative mass inside the quantum dot and positive outside it or vice versa) gives rise to a new spectral curve of  edge states. All these cases have been analyzed in detail along this paper.
\end{abstract}

\section{Introduction}

In the specialized literature there is a wide variety of works on position depending positive mass. On the other hand, the connection of edge states \cite{Cayssol13,Shen17,Lado15,Lado22} versus opposite sign mass \cite{Myoung21,Belokda22,sari22,belokda23}. The present work focuses on showing the influence that different (constant) effective masses inside and outside an electric quantum dot can have on the spectrum  of two-dimensional Dirac material. We pay special attention to the case in which the masses have different signs in these two regions \cite{belokda23,Downing19}. Our plan is to provide a general perspective based on the determination of the solutions in each constant mass domain.
The solutions in each domain depend on the parameters $\varepsilon$ (energy), $\mu$ (mass) and $v$ (potential depth). We will work out these dependences on these parameters.
Therefore,  we consider the energy ($\varepsilon$) versus the mass $\mu$ (in a $\varepsilon$--$\mu$ plane) and the energy versus the potential depth $v$ (in a $\varepsilon$--$v$ plane).  
Each two parameter planes are divided into regions. Then we select the right solutions of each parameter region with correct behavior in asymptotic spatial boundaries: when the radius goes to 0, $\infty$ and to the radius of the quantum dot.

Let us mention some of the most interesting results we will obtain: 
(a) there are ``critical points'' that consist in the end points of the spectral curves, which we will characterize in detail. These critical points determine the values where the system captures or looses a bound states. They are associated to collapses due to relativistic character of the Dirac equation. (b) the spectrum is quite sensitive to changes in the effective masses in the inside and outside regions, 
and  (c) the case of opposite masses give rise to the so-called ``edge states''.

The type of quantum dots that we study here have a wide field of applications in quantum optics \cite{recher10,Zwiller04,Beveratos14}, but it should be noted that there are studies on other ways of confining using electric fields, such as those discussed in \cite{Downing19}, or in nanoribbons \cite{Peeters21,Vit22}, or also in quantum dots that are not two-dimensional but spherical \cite{Bilynskyi22}.

The structure of the work is the following. In Section~\ref{direq} the Dirac equation for this problem is presented and the spectral problem is defined. Section~\ref{cases} details the solutions for all possible cases that may arise. The work ends with the conclusions that are presented in Section~\ref{conclu}. The details of the computations, for the interested reader, are presented in Appendix~\ref{Appendix}.

\section{Two-dimensional Dirac particle in an electric  dot}\label{direq}

Consider grphene, or a Dirac planar material, where in the low energy regime near the K Dirac points, the electronic states are described by a Dirac-Weyl equation  \cite{Cayssol13,Shen17} .  In our case, we wil consider the material subject to an external electrostatic potential  $V({\bf r})$ is:
\begin{equation}\label{2DDH}
H = v_F \, {\boldsymbol \sigma}\cdot{\bf p} + m \, v_F^2\, \sigma_z+  e\, V({\bf r}), 
\qquad  {\bf r}=(x,y)\in\mathbb{R}^2 ,
\end{equation}
where ${\boldsymbol \sigma} = (\sigma_x,\sigma_y)$ and $\sigma_z$ are Pauli matrices,
${\bf p} = (p_x,p_y) = -i\hbar(\partial_x,\partial_y)$ are  the momentum operators, and
$v_F$ is the Fermi velocity of the material. 
In the present work we are  going to consider the properties of the charged particles (electrons) inside an electric quantum dot
such that the electric potential is a constant two-dimensional radial well of the form
\begin{equation}\label{potentialwell}
V({\bf r}) = \left\{\begin{array}{cl}
V_{\rm i},  & r<R,
\\[1ex]
V_{\rm o},  & r>R,
\end{array}\right.
\end{equation}
but, at the same time, we also assume that the mass of the particle can be different inside (${\rm i}$) and outside (${\rm o}$) of this well:
\begin{equation}\label{masslwell}
m = \left\{\begin{array}{cl}
m_{\rm i},  & r<R,
\\[1ex]
m_{\rm o},  & r>R\,.
\end{array}\right.
\end{equation}
Therefore, the 2D Dirac Hamiltonian \eqref{2DDH} also has two different expressions, one  inside
and one outside 
the electric quantum dot:
\begin{equation}\label{2DDHtwocomponents}
H = \left\{\begin{array}{cl}
v_F \, {\boldsymbol \sigma}\cdot{\bf p} + m_{\rm i} \,v_F^2\, \sigma_z+  e\, V_{\rm i},  & r<R,
\\[1.1ex]
v_F \, {\boldsymbol \sigma}\cdot{\bf p} + m_{\rm o}\, v_F^2\, \sigma_z+  e\, V_{\rm o},  & r>R.
\end{array}\right.
\end{equation}
From the approach made so far,  radial symmetry is obvious, and therefore  it is natural to use polar  coordinates $(r,\theta)$ to  
separate variables in the time-independent Dirac equation. In the present case, the Hamiltonian commutes with the total angular momentum  operator $J_z$ about the $z$--axis, defined by
\begin{equation}
J_z = L_z + \Sigma,\quad {\rm with}\quad L_z = -i\hbar \partial_\theta \quad {\rm and}\quad
\Sigma = \frac 12\hbar \sigma_z \,.
\end{equation} 
Therefore, we can look for the spinors $\Phi(r,\theta)$ that  are simultaneous  eigenfunctions of $H$ and $J_z$:
\begin{equation}\label{j}
H \Phi(r,\theta) = E\Phi(r,\theta),\qquad\qquad J_z\Phi(r,\theta) = j\hbar \Phi(r,\theta).
\end{equation}
The second of the equations  in (\ref{j}) leads to eigenfunctions  of the form
\begin{equation}\label{phi}
\Phi_\ell (r,\theta) = \left(\begin{array}{c}
\phi_1(r)\ e^{i \ell \theta}
\\[1.ex]
i \phi_2(r)\ e^{i (\ell+1) \theta}\end{array}\right) , \qquad \ell=  j-\frac12 , \quad \ell= 0,\pm1,\dots,
\end{equation}
where, for convenience, the imaginary unit has been introduced in the second component. 
The normalization of  the spinors (\ref{phi}) is defined in the usual way:
\begin{equation}
\int_{\mathbb{R}^2} \Phi_\ell(r,\theta)^\dag\ \Phi_\ell(r,\theta)\, r\, dr\, d\theta=
2\pi \int_0^\infty \left( | \phi_1(r)|^2+| \phi_2(r)|^2\right)\, r\, dr =1.
\label{normalization}
\end{equation}


To find the differential equations that must satisfy the $\phi_1(r), \phi_2(r)$ components of the eigenspinors, we plug (\ref{phi}) into the eigenvalue equation of $H$ expressed in polar coordinates. 
So, we get two reduced equations in the variable $r$:
\begin{eqnarray}\label{spinorialsystem}
\left(\begin{array}{cc}
0 & -i A^-
\\[1.ex]
iA^+ & 0 \end{array}\right)  
\left(\begin{array}{c}
\phi_{1,k}(r)
\\[1.ex]
i \phi_{2,k}(r)\end{array}\right) =
\left(\begin{array}{cc}
E{-}{ e}V_k{-}m_k v_F^2 & 0
\\[1.ex]
0 & E{-}{ e}V_k{+}m_k v_F^2 \end{array}\right)  
\left(\begin{array}{c}
\phi_{1,k}(r)
\\[1.ex]
i \phi_{2,k}(r)\end{array}\right),
\end{eqnarray}
where $k={\rm i}, {\rm o}$ indicates the interior or exterior regions of the quantum dot, and the operators $A^\pm$ are
\begin{equation}\label{aminusplus}
A^-= \hbar v_F\left(\partial_r +\frac{\ell+1}{r}\right),\quad  
A^+= \hbar v_F\left(-\partial_r +\frac{\ell}{r}\right),\quad \ell\in \mathbb{Z}.
\end{equation}
We now redefine   the independent variable and the relevant physical  parameters to simplify the problem:
\begin{equation}\label{newvar}
\rho=\frac{r}{R},\quad 
\varepsilon = \frac{E R}{\hbar v_F},\quad 
 v_k=  \frac{e V_k R}{\hbar v_F},\quad 
\mu_k =  \frac{m_k v_F R}{\hbar },\qquad k={\rm i}, {\rm o},
\end{equation}
where the dimensionless potentials $v_k$ and  masses $\mu_k$ are constants. 
Then,  the system \eqref{spinorialsystem} becomes the following pair of coupled differential equations:
\begin{equation}\label{phis}
\left\{
\begin{array}{r}
\displaystyle 
\frac{d\phi_{2,k} (\rho)}{d\rho} +\frac{\ell+1}{\rho}\ \phi_{2,k} (\rho)= (\varepsilon -v_{k} -\mu_{k}) \ \phi_{1,k} (\rho), \\[2ex] 
\displaystyle 
- \frac{d\phi_{1,k} (\rho)}{d\rho} + \frac{\ell}{\rho}\ \phi_{1,k} (\rho) =  (\varepsilon -v_{k} +\mu_{k}) \  \phi_{2,k} (\rho),
\end{array}\right.
\end{equation}
where the subindex is $k={\rm i}$ for the inner region of the dot ($0\leq\rho<1$) or $k={\rm o}$ for the exterior region ($\rho>1$).
Since only the difference $v_{\rm i}-v_{\rm o}$ is important, from now on we will take $v_{\rm o}=0$ (for the potential to be zero at infinity), and for simplicity we will write $v_i\equiv v$ inside the quantum dot.

The differential equations satisfied by the first components $\phi_{1,k}$ of the spinors are obtained from (\ref{phis}):\begin{equation}\label{phi_1}
\frac{d^2\phi_{1,k} (\rho)}{d\rho^2}+\frac1\rho\, \frac{d\phi_{1,k} (\rho)}{d\rho}- \left( \mu^2_{k}-(\varepsilon-v_k)^2 + \frac{\ell^2}{\rho^2} \right)\phi_{ 1,k} (\rho) =0, \quad k={\rm i},{\rm o}\, .
\end{equation}
Once these first components of the spinor are known, the second ones are easily obtained as
\begin{equation}\label{phi_2}
\phi_{2,k} (\rho)= \frac1{\varepsilon -v_{k} +\mu_{k}} \left[   - \frac{d\phi_{1,k} (\rho)}{d\rho} + \frac{\ell}{\rho}\ \phi_{1,k} (\rho)   \right]  , \quad k={\rm i},{\rm o}\, .
\end{equation}
The   previous expressions may not be valid when   $\varepsilon =v_{k} \pm \mu_{k}$ (with $ k={\rm i},{\rm o}$). They are limit situations that we will analyze separately. These situations have special solutions called either ``critical'' or ``extreme'', as we will specify later.

In order to  simplify the notation used, we will denote the outer mass as $\mu_{\rm o}\equiv \mu>0$, since it will be taken as the reference mass, and the inner mass as $\mu_{\rm i}=\lambda \mu$, where $\lambda$ is a real constant (which can be positive or negative) to be specified.

Given that in the differential equations \eqref{phi_1}--\eqref{phi_2} there are a total of four parameters  ($\varepsilon, v, \mu, \lambda$), and the solutions depend on them, to simplify the analysis, it is convenient to consider the plane $\varepsilon$--$\mu$ and there to distinguish several regions, more precisely:
\begin{itemize}
\item
The so-called {\it region outside} the quantum dot ($RO$), defined by the inequality $\mu^2 > \varepsilon^2$, which are the conditions that must be met for the bound state wave functions to be square integrable outside the quantum dot. The limit cases 
$\varepsilon = \mu$ and $\varepsilon=-\mu$ are called, respectively,  critical and supercritical lines \cite{Hall,Negro22}.
\item
For the solutions inside the quantum dot there are three different possibilities that must be analyzed separately: either 
$ (\lambda \mu)^2- (\varepsilon -v)^2<0$, which determines a part of the $\varepsilon$--$\mu$ plane that we will call $RI_1$, or $ (\lambda \mu)^2- (\varepsilon -v)^2>0$, which determines a different part of the $\varepsilon$--$\mu$ plane that we will denote as $RI_2$, or finally the so-called {\it extreme case} when the equality $ (\lambda \mu)^2- (\varepsilon -v)^2=0$ is fulfilled.
\end{itemize}

\begin{figure}[h]
\begin{center}
  \includegraphics[width=0.45\textwidth]{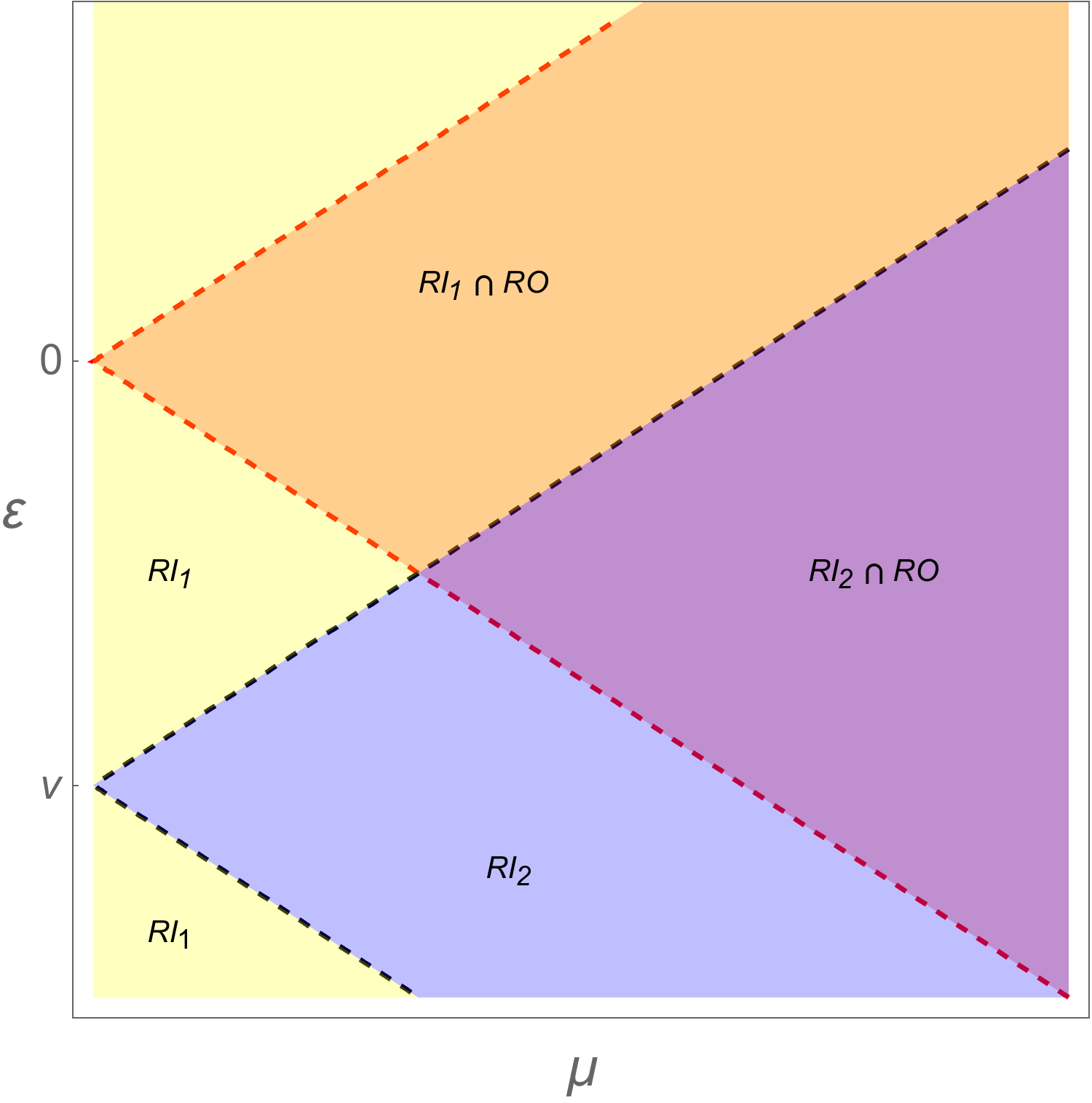}
     \caption{\small 
    For a fixed negative value of the potential inside the quantum dot ($v<0$), plot of the regions $RO$, $RI_1$ and $RI_2$, where bound states solutions may exist. The two black dashed  lines correspond to the {\it extreme lines} and the two red dashed rays are the {\it critical} ($\varepsilon=+\mu$) and {\it supercritical} ($\varepsilon=-\mu$) lines. For simplicity, we have used $\lambda=1$.
}
  \label{ROI12}
  \end{center}
\end{figure}

Remark that while this regions have been defined based on the sign of the constant term of  (\ref{phi_1}), the original problem (\ref{phis}) is continuous in the parameters ($\varepsilon, v, \mu, \lambda$). Thus, while we are going to showcase next the solutions inside each parameter region as independently, this solutions are continuous and represent the same physical entity. 
A schematic of what has just been exposed can be seen in Figure~\ref{ROI12}, where for a value $v<0$ the three regions $RO$, $RI_1$ and $RI_2$ and their intersections are clearly indicated by the colors.
Next, we are going to analyze in a mathematically rigorous way what happens in each of the cases that have been commented.

\section{Analysis of the bound states depending on the mass} \label{cases}

 We are going to qualitatively analyze the various types of bound states that, can exist attending to the different $\lambda$ values. For each $\lambda$ value we plot the energy values that arise from the various matching equations, whose explicit expressions are contained in the Appendix~\ref{Appendix}.
\newline
The figures of the  plots within each of the following subsections have the same structure:
\begin{itemize}
    \item[] The left graph represents the different bound state energy levels, as functions of $\mu$, that arise for a fixed potential  value $v$.
    \item[] The right graph represents the different bound state energy levels, as functions of the potential $v$, that arise for a fixed mass $\mu$.
      \item[] We introduce a green line to both graphs in order to add some connection between them, as this line represents the the system with the same values of v and $\mu$, and therefore have the same spectrum in the left and in the right figures.. Thus this energy lines represent the same problem in both graphs.
    \end{itemize}
    
\subsection{Positive masses $\lambda>0$}
In this section we will discuss how the $\lambda$ parameter modifies the region $RI_1\cap RO$, as well as the consequences of this region changes. Since,  as can be seen in the Appendix~\ref{Appendix}, the solutions admitted by the system for $\lambda>0$ must lay inside the region $RI_1\cap RO$.
As we have seen, for  positive masses, $\lambda>0$,

\subsubsection{Equal masses  $\mu= \mu_{\rm i}=\mu_{\rm o}$ (${\lambda =1}$)}

Here we simply have the problem of constant mass which
has been considered previously in \cite{Negro22} and references therein. The representation of energy values obtained is the following:


\begin{figure}[h!]
\begin{center}
    \includegraphics[width=0.39\textwidth]{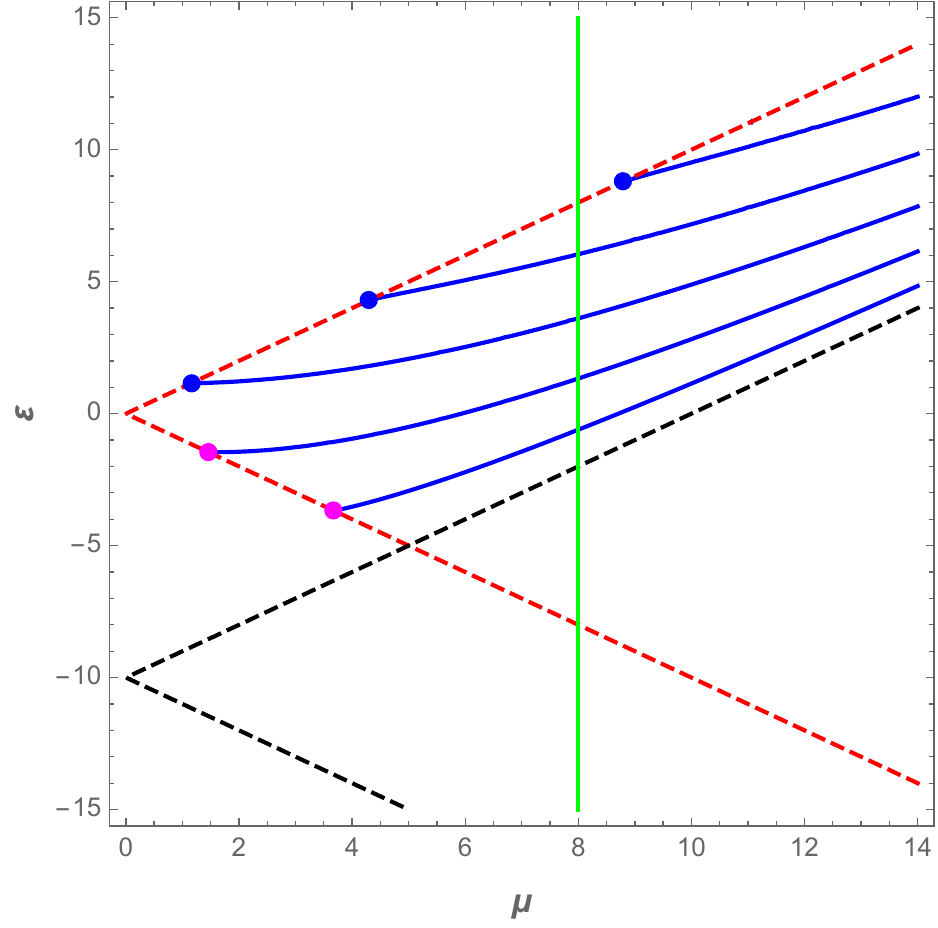}\quad
    \includegraphics[width=0.38\textwidth]{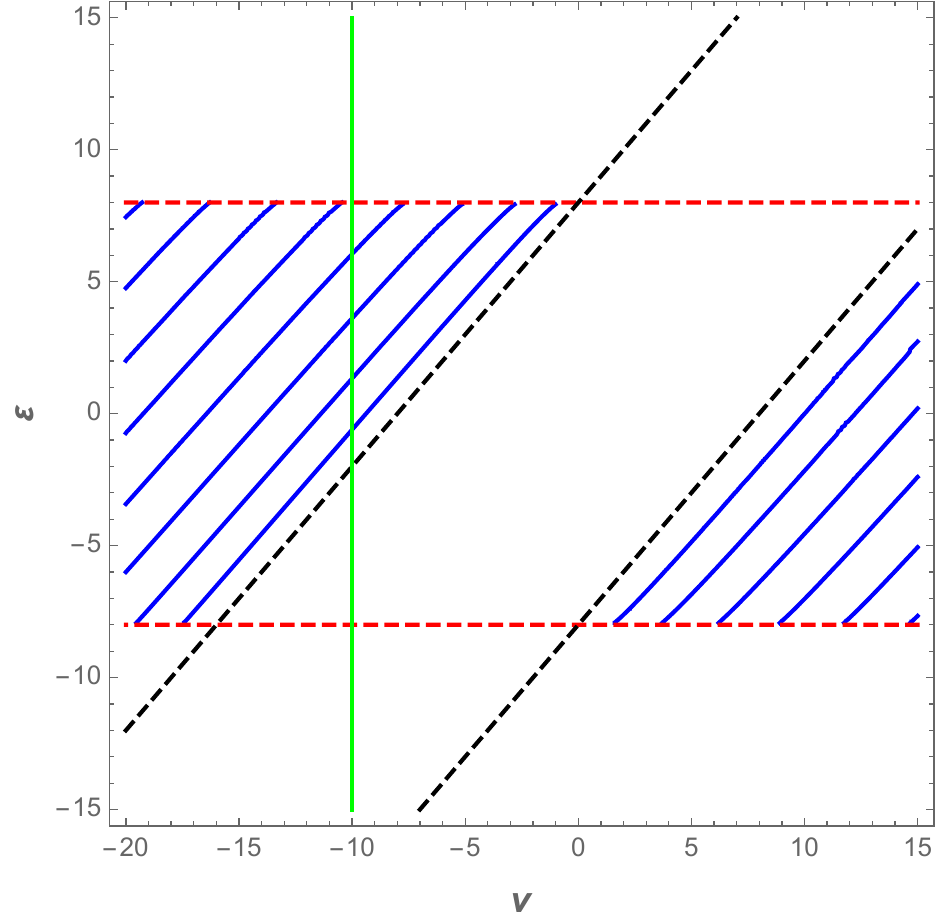}
 \caption{\small 
(Left) Solutions (in blue lines) of the secular equation, for $\ell=2$,
when both masses are equal and positive, on the plane $\varepsilon$-$\mu$ for a constant value of $v =-10$. In this case they belong to the intersection regions $RI_1{\cap}RO$ of 
Fig.~\ref{fsectors}. 
}
  \label{fsectors1}
  \end{center}
\end{figure}
\newpage 
The critical in blue  and  supercritical in magenta points are represented on the lines $\epsilon = \pm \mu$. (Right) Here, the  blue lines represent the discrete energies of the quantum dot as functions of the inner potential $v$ (keeping the mass constant $\mu=8$) also for equal inner and outer masses. These spectral lines of energy levels $\varepsilon$ are bounded by the dashing red lines of values $\pm \mu$.  The green line at $v =-10$ gives the same spectrum as the green line at $\mu=8$ in the $\varepsilon$-$\mu$ of the l.h.s.

\subsubsection{Inner mass smaller than outer mass ${ \mu_{\rm i}<\mu_{\rm o}}$ (${|\lambda| <1}$)}
\begin{figure}[h!]
\begin{center}
  \includegraphics[width=0.4\textwidth]{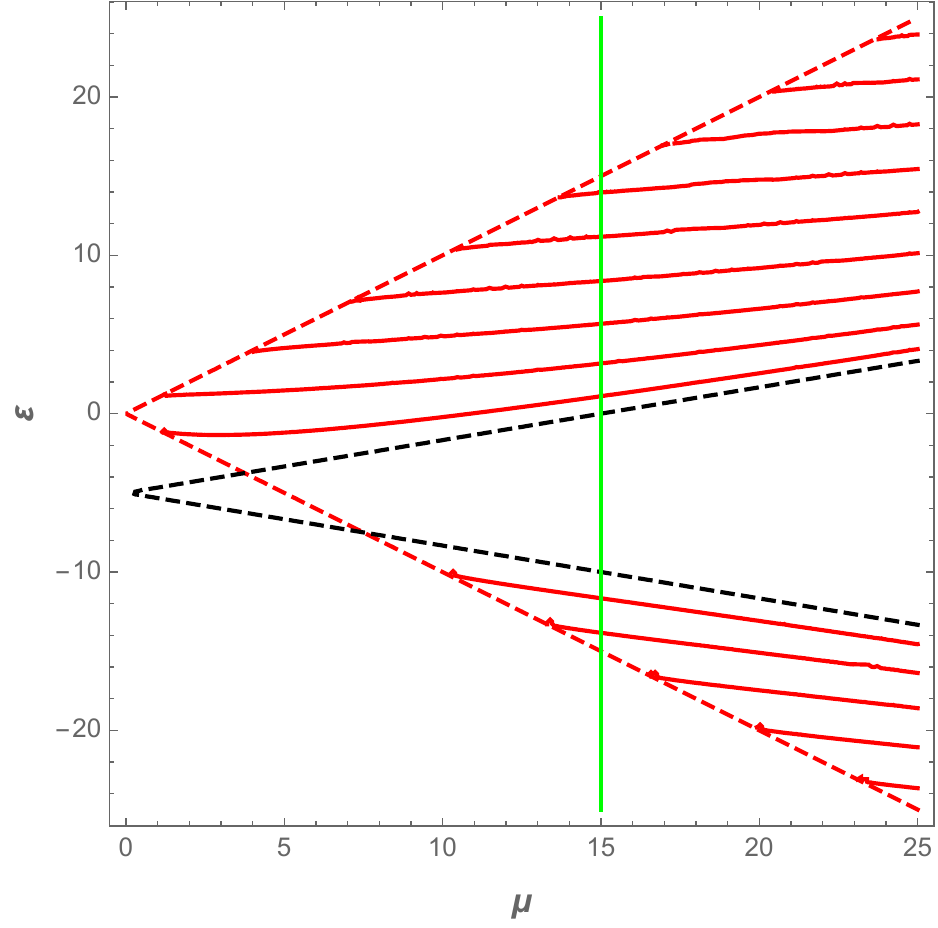}\qquad
  \includegraphics[width=0.4\textwidth]{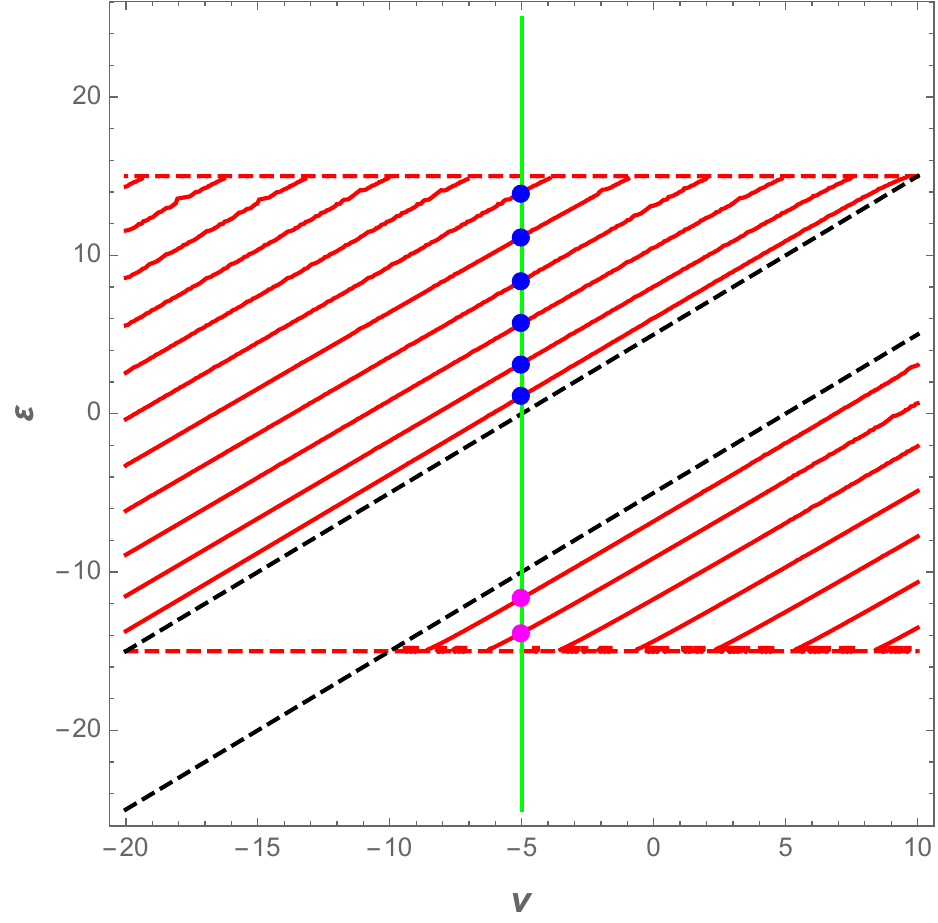}
 \caption{\small
 (Left) Energy eigenvalues, for a fixed value $\ell=1$, as a function of the outer mass $\mu$, where the inner mass is lighter than the outer, $\lambda=\frac{1}{3}$, for $v_{\rm i}=-5$. (Right) A plot $\varepsilon$-$v$, considering a fixed value of outer mass $\mu=15$. We can appreciate the gap between the upper and lower levels due to the inner cone in dashing black lines.}
  \label{fig4}
  \end{center}
\end{figure}

As shown in the next figure, one key
feature of this case is that the region $RI_1 \cap RO$ of possible values, for 
inner lighter mass, has been increased, so that there will be many more
bound states (spectral curves) as can be appreciated in Fig.~\ref{fig4}.

 More importantly, inside that spectrum
there is a gap between the lower and the higher energy levels, $\Delta \varepsilon>2|\lambda|\mu$. Therefore, there are two types of spectrum points belonging to these
subregions. The upper points (in blue in Fig.~\ref{fig4} right) come from the capture of bound states of the positive continuous part of the spectra. This determines what are the ground and excited states of this part. However, the points of the lower subregion (in magenta in Fig.~\ref{fig4} right) are interpreted as coming from the negative continuous spectra, which determines the corresponding ordering of ground and excited states. Both types of states belong to the discrete spectrum of the Hamiltonian and they are orthogonal. 

For instance, the system described in Fig.~\ref{fig4} (determined by the cuts of the green vertical line) consist of two lower bound states with energy $\varepsilon_0^-=-11.6592$ and $\varepsilon_1^-=-13.8407$; while the points of the upper spectrum start with the values $\varepsilon_0^+=1.09801$ and $\varepsilon_1^+=3.16259$, \dots, $\varepsilon_6^+=13.94916$, where $\varepsilon_i^+$ designs the  i--excited state of the upper spectrum, and $\varepsilon_k^+$ designs the  i--excited state of the lower spectrum.

\noindent
\subsubsection{Inner mass greater than outer mass 
$ {\mu_{\rm i}>\mu_{\rm o}}$ ($ {|\lambda| >1}$)}

\begin{figure}[h]
\begin{center}
  \includegraphics[width=0.4\textwidth]{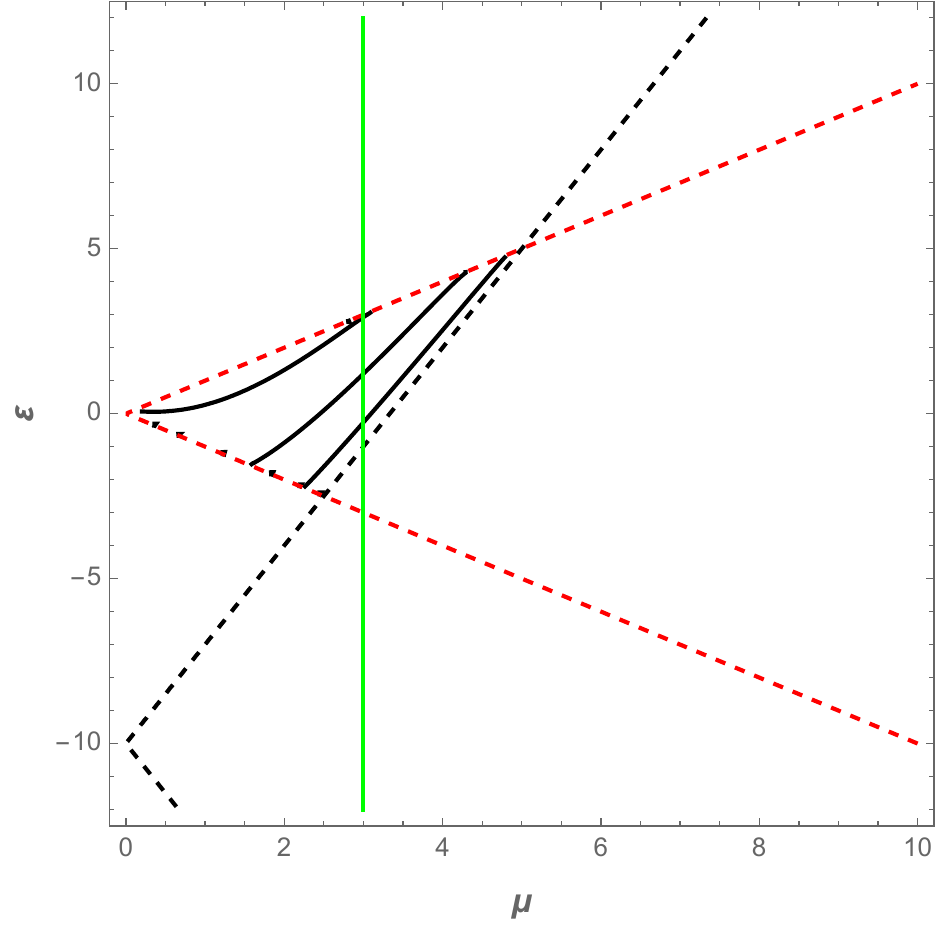}\qquad
  \includegraphics[width=0.4\textwidth]{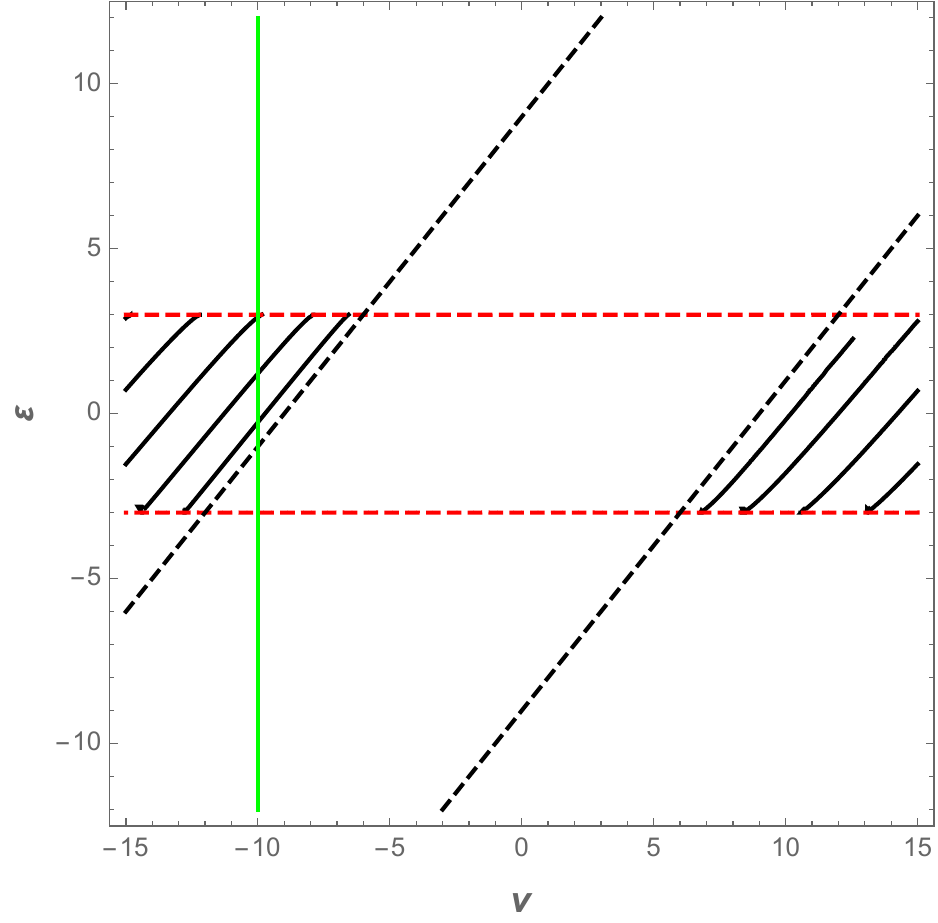}
 \caption{\small
(Left) Energy values $\varepsilon$, for $\ell=1$, as a function of the outer mass $\mu_{\rm o}:=\mu$, where $\mu_{\rm i}= 3\mu $ for $v = -10$. (Right) Graphic of the energy levels $\varepsilon$ versus the potential depth  $v$, for $\mu=3$. 
The green line corresponds to $v=-10$ (right) and $\mu=3$ (left).
 }
  \label{fig5}
  \end{center}
\end{figure}

In this case, the behavior of the spectrum is opposite to that of the previous case. The region of energy curves is reduced,  it becomes finite as can be seen in Fig.~\ref{fig5} (left) and the number of bound states becomes very low.  In order to confine states it is necessary  very intense potentials, see Fig~\ref{fig5}(right).

\subsection{Masses of opposite sign $\lambda<0$ and edge states}
 While the sign of $\lambda$ does not lead to any changes in the energy regions (Fig.~\ref{fsectors}), this mass inversion leads to important consequences. As can bee seen in the Appendix~\ref{Appendix}, the system may now admit solutions inside the region $RI_2\cap RO$. This special type of solutions are a kind of ground states which are called ``edge states'' \cite{Cayssol13,Shen17,Asorey13,Dolcetto}. In this section we will discuss the problem of confinement that arise from this new kind of solutions.



\subsubsection{Equal absolute masses ${ \mu_{\rm i}=-\mu}$ (${\lambda =-1}$)
\label{momim}}
\begin{figure}[h]
\begin{center}
  \includegraphics[width=0.4\textwidth]{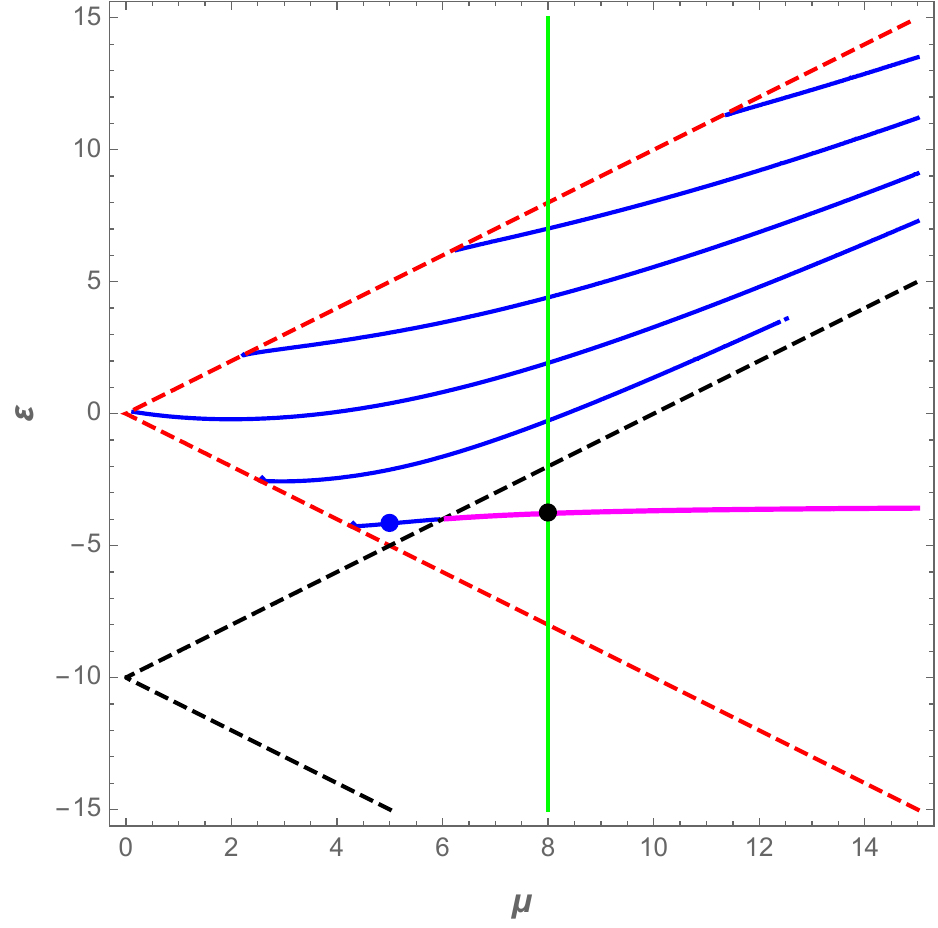}\quad
   \includegraphics[width=0.4\textwidth]{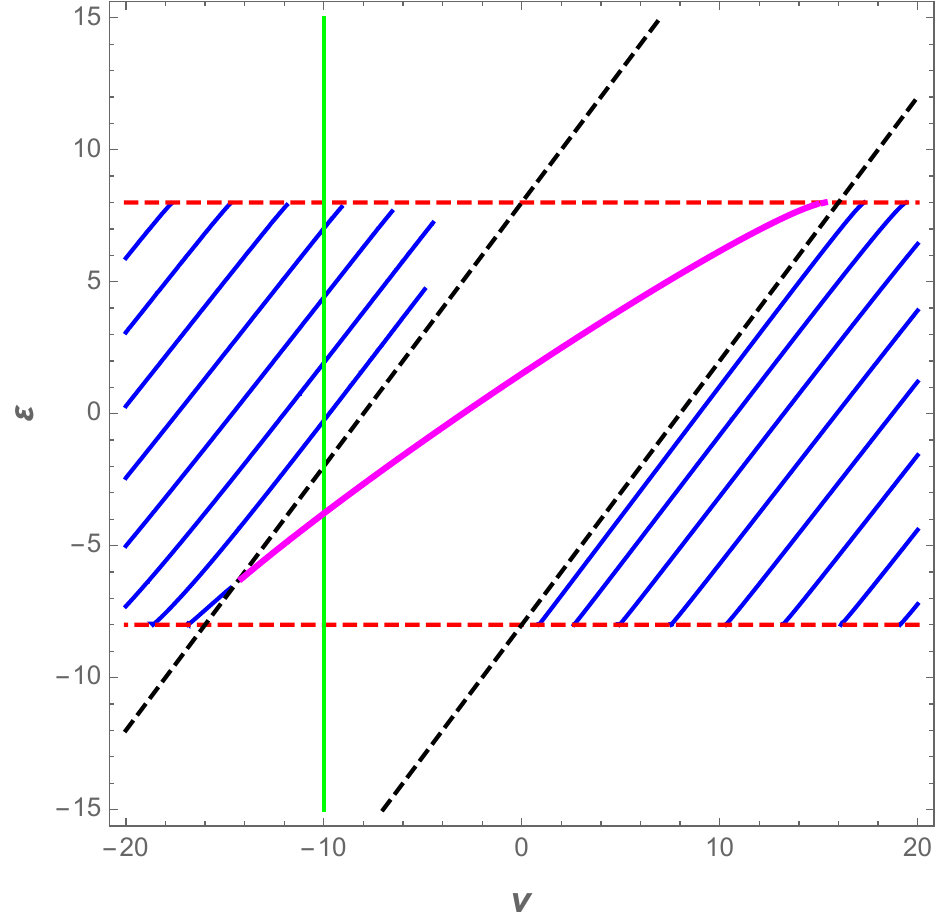}
 \caption{\small
Graphics for Subsection~\ref{momim}  with negative $\mu_{\rm i}= -\mu$. (Left) Plot of the spectrum in the plane $\varepsilon$-$\mu$ with $v=-10$. (Right) Energy values as a function of the potential intensity $v$ and where the mass is fixed $\mu_{\rm i}=-8$. The spectral curve of the edge states is in magenta, it is the prolongation of a piece of curve of extended states (in blue). Two dots are shown, one in the edge part in red and another, in blue, on the extended part of the same curve.}
  \label{fig6}
  \end{center}
\end{figure}

As can be seen in Fig.~\ref{fig6} (left) the new energy level corresponding to the edge state  is the spectral curve included in the intersection of the two cones in dashing lines, that is, in the region $RI_2\cap RO$. 
  \begin{figure}[h!]
\begin{center}
  \includegraphics[width=0.5\textwidth]{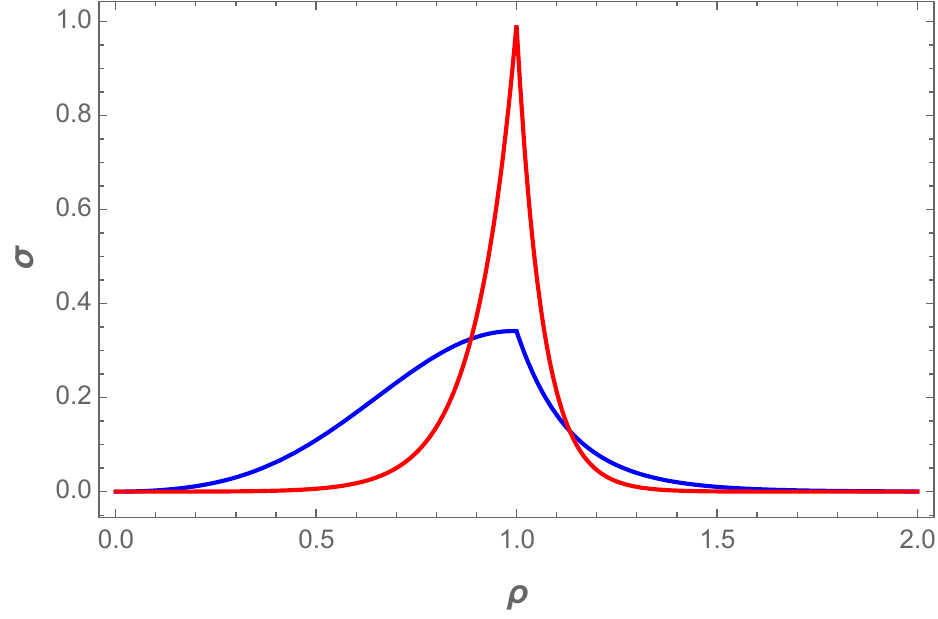} 
 \caption{\small
Plot of the probability density $\sigma$ of states of the spectral points belonging to the ground state curve shown in Fig.~\ref{fig6} (left).
  The {\bf edge state} (in red) with parameters: $\varepsilon = -3.78$, $ \mu_{\rm i} = -8$, $\mu = 8$, $ v  = -10$, $ \ell =
  1$.  The (blue) standard {\bf bulk ground state}  with parameters: $\varepsilon = -4.16$, $ \mu_{\rm i} = -5$, $ \mu = 5$, $ v  = -10$, $\ell = 1$.}
  \label{probability1}
  \end{center}
\end{figure}

We have selected two
points in the ground state curve of Fig.~\ref{fig6} (left): a) the first
in blue, belongs to the region $RI_1\cap RO$, which is not an edge state; and
b) a second point in red on the spectral edge curve (cyan) that belongs to the region $RI_2\cap RO$, so it represents an edge state.  
 
We have computed their corresponding wave functions and
their resulting probability densities are displayed in Fig.~\ref{probability1}.
It is shown that both describe ground states; the blue one is for an
extended or bulk state, but the red one, on a cyan edge curve, is for an edge state near the border $\rho=1$
of the dot.

This edge state solution  can
also be appreciated in the plot of the $\varepsilon$--$v $ plane of Fig.~\ref{fig6} (right). This spectral curve starts at the
upper red dashing line, so it is clear from the figure that it corresponds to ground state level.
The new spectral line is distinguished from the rest because  it belongs to the region $RO\cap RI_2$,  a region which was empty  of spectral lines for the equal  sign mass case.

\subsubsection{Absolute inner mass smaller (${ |\mu_{\rm i}|<\mu_{\rm o}}$) and greater (${ |\mu_{\rm i}|>\mu_{\rm o}}$)  
mass
}

\begin{figure}[h!]
\begin{center}
  \includegraphics[width=0.4\textwidth]{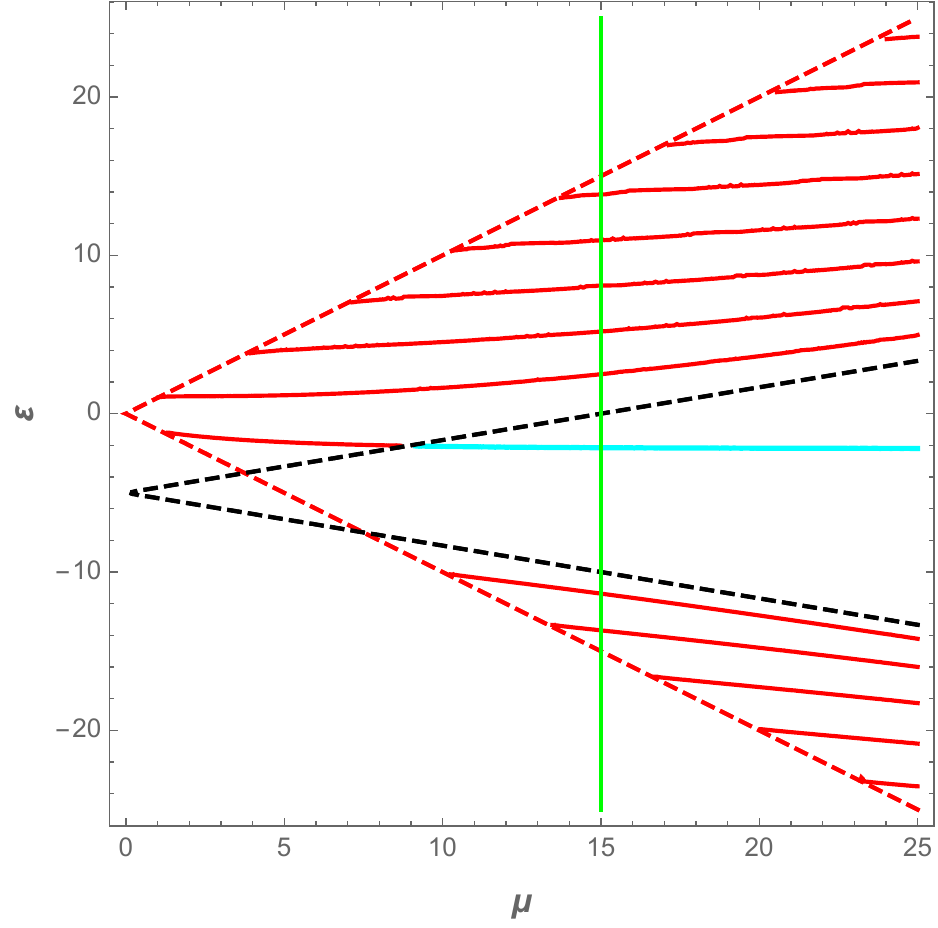}\qquad
   \includegraphics[width=0.4\textwidth]{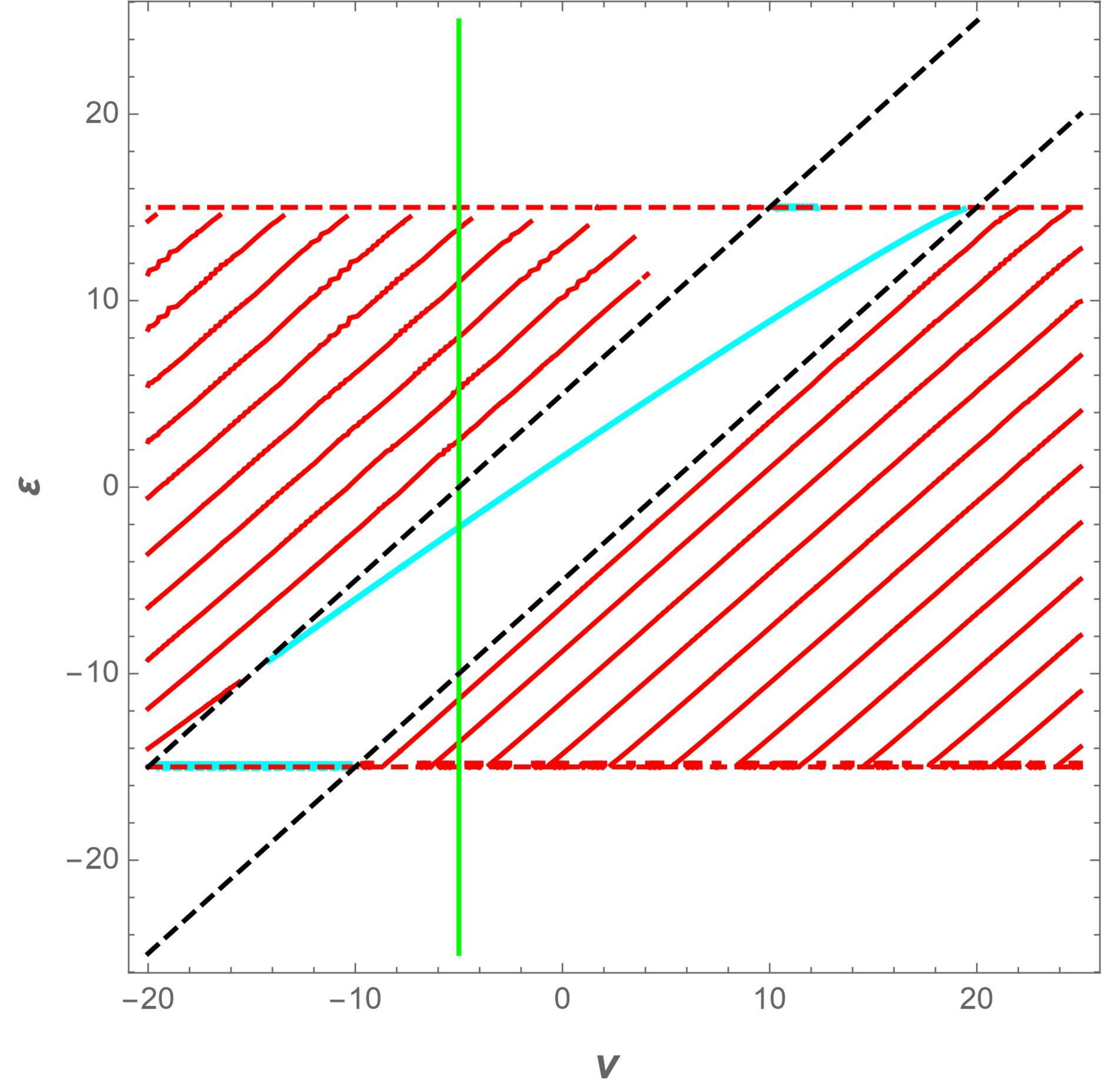}
 \caption{\small
Graphics for the case $\ell=1$, ${ |\mu_{\rm i}|<\mu_{\rm o}}$ with negative $\mu_{\rm i}$. (Left) Plot of the spectrum in the plane $\varepsilon$-$\mu$ with $v=-5$. (Right) Energy values as a function of the potential intensity $|v|$ and where the mass is fixed $\mu_{\rm i}=-5$.}
  \label{fig8}
  \end{center}
\end{figure}

\begin{figure}[h!]
\begin{center}
  \includegraphics[width=0.4\textwidth]{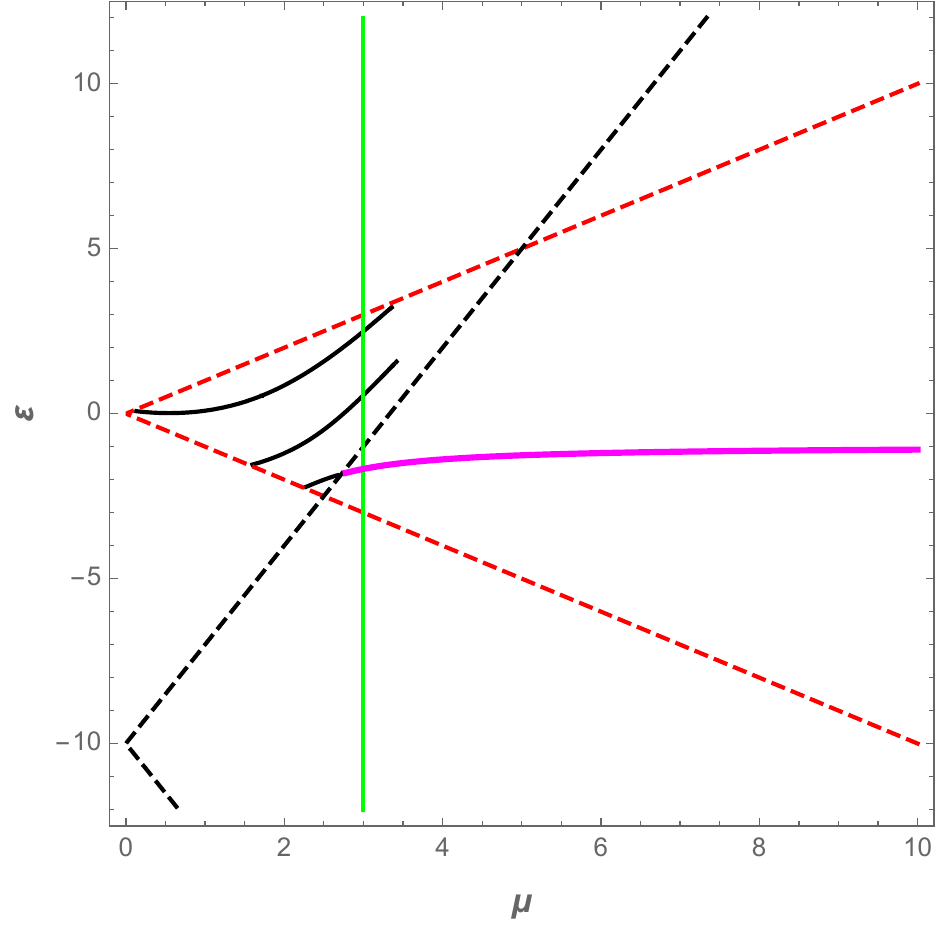}\qquad
   \includegraphics[width=0.4\textwidth]{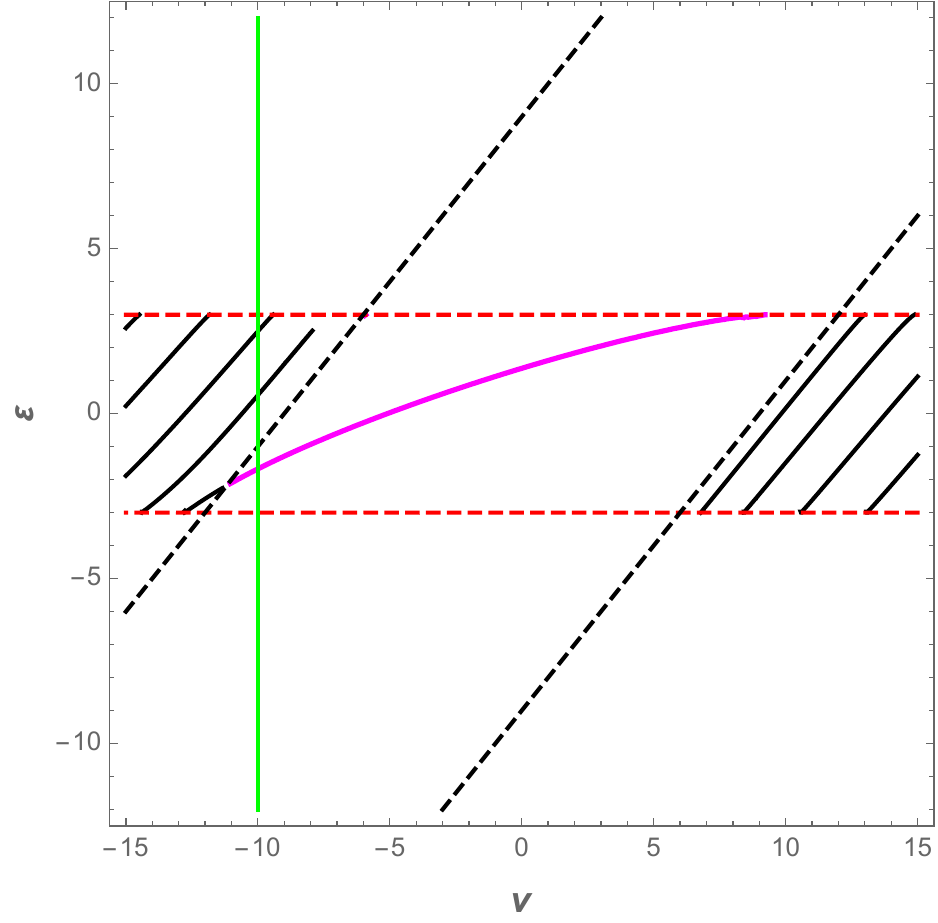}
 \caption{
 \small
Graphics for the  case ${ |\mu_{\rm i}|>\mu_{\rm o}}$ with negative $\mu_{\rm i}$. (Left) Plot of the spectrum in the plane $\varepsilon$-$\mu$ with $v=-10$. (Right) Energy values as a function of the potential intensity $|v|$ and where the mass is fixed $\mu_{\rm i}=-9$.}
  \label{fig9}
  \end{center}
\end{figure}

Similar considerations with respect to the spectral curves apply in these two remaining cases. There is a new solution in the region $RI_2\cap RO$, corresponding
edge states  describing ground states.
The graphics of these cases are given in Figs.~\ref{fig8}-\ref{fig9}, for  
${ |\mu_{\rm i}|<\mu_{\rm o}}$ and ${ |\mu_{\rm i}|>\mu_{\rm o}}$, respectively.   
\newpage

\section{Conclusions}\label{conclu} 
This paper was devoted to discuss in detail  the influence that the different masses, outside and inside, of an  electric quantum dot have on its spectral properties.
We have considered two  qualitative situations: the case where inner and outer mass have the same or if they have opposite  sign. In this work we have made use of two types of diagrams (energy--mass or $\epsilon$--$\mu$, and
energy--potential depth or $\epsilon$--$v$)  which are complementary and allows us to describe in full detail de main properties of the spectrum.
\vspace{2mm}
The analysis of the equal sign situation is divided into three subcases whose  main features are as follow:
\begin{itemize}
    \item[(a)] $|\lambda|=1$ Case. Identic masses. This is the base case in our study and corresponds to \cite{Negro22}. 
    \item[(b)] $|\lambda|<1$ Case. Inner mass lower than exterior mass. 
  Under this conditions, the spectrum of the dot is more dense and it splits in two parts separated by a gap. This splitting appears for parameter values that satisfy $v^2<(|\lambda|-1)^2\mu^2$ and the energy gap satisfy $\Delta \varepsilon>2|\lambda|\mu$.  
  
    \item[(c)] $|\lambda|>1$ Case. Inner mass greater than exterior mass. 
    Under this conditions, the spectrum of the dot becomes less dense. Contrary to the prior case, the system does not admit bound states for parameter values $v^2<(|\lambda|-1)^2\mu^2$.
\end{itemize}
On the other hand, the same subcases have been considered for the opposite sign case. Where we obtain the same results as the case prior, plus a new spectral curve inside the energy region $RI_2\cap RO$. The main result is the presence of a new spectral curve that is interpreted as made up of spectral points of edge states, which describe a ground state. We have plotted the corresponding wave function and have commented on its role inside the total spectrum of the system.

 These results can be implemented in two-dimensional Dirac materials such as graphene as described in \cite{Asmar}.  
These new materials will act as a candidate for quantum computing hardware construction, since the edge states allow for an easy and  compact way of electron confinement that ultimately corresponds to an optimum qubit \cite{Lado22,Titov}.

 \section*{Acknowledgments}

This research was supported by the Q-CAYLE project, funded by the European Union-Next Generation UE/MICIU/Plan de Recuperacion, Transformacion y Resiliencia/Junta de Castilla y Leon (PRTRC17.11), and also by project PID2023-148409NB-I00, funded by MICIU/AEI/10.13039/501100011033. Financial support of the Department of Education of the Junta de Castilla y Leon and FEDER Funds is also gratefully acknowledged (Reference: CLU-2023-1-05).

\section{Appendix}\label{Appendix}
This section contains the explicit analysis of the bound states of the electric quantum dot given by differential equations \eqref{phi_1}--\eqref{phi_2}.
\subsection{Solutions outside the quantum dot ($1\leq\rho$)}

For the bound-state wave functions to be square integrable, they must be such that outside the quantum dot they have an appropriate asymptotic behavior, such as exponential decay. Since we have taken $v_{\rm o}=0$, for this to happen, we deduce from \eqref{phi_1} that we must have
$\mu^2\geq \varepsilon^2$, that is, 
\begin{equation}\label{-muoepsilonmuo}
-\mu\leq \varepsilon\leq \mu,
\end{equation}
and therefore the value of the mass outside the quantum dot constrains the possible energy eigenvalues of the bound states.

As we are going to study in the plane $\mu$--$\varepsilon$ the possible existence of eigenvalues $\varepsilon$ as a function of the mass $\mu$, keeping the rest of the parameters fixed, we are forced to consider separately two possibilities that give rise to different types of solutions.

\begin{itemize}
\item[(i)]{Solutions in the parameter region  $RO$:  $ \mu^2> \varepsilon^2$.}

The acceptable values of $\varepsilon$ must lie within the region between the two lines $\varepsilon=\pm \mu$, which lie in the half-plane $\mu >0$. This is what was called the $RO$ region (or region-out) in the initial comment, and is shown in Figure~\ref{ROI12}.
To be more precise, in this region $RO$ the equation ~(\ref{phi_1}) is a modified Bessel equation \cite{abramowitz}, and therefore it is well known that solutions of \eqref{phi_1}--\eqref{phi_2} and \eqref{phi} that are physically acceptable turn out to be
\begin{equation}\label{phiout}
\Phi_{\ell,\rm o}(\rho,{\theta}) = a_\ell \left(
\begin{array}{c}
\sqrt{\mu+\varepsilon}K_\ell( \rho\sqrt{\mu^2- \varepsilon^2}\,) \  e^{i \ell \theta}
\\[1.5ex]
\displaystyle i \, \sqrt{\mu- \varepsilon} \, 
K_{\ell+1} ( \rho\sqrt{\mu^2- \varepsilon^2}\,) \ e^{i (\ell+1) \theta}
\end{array}\right),
\qquad \ell\in \mathbb{Z},
\end{equation}
\newpage
where  $K_\nu(z)$ is the modified Bessel function of the second kind and order $\nu$ \cite{abramowitz}, and $a_\ell$ is a normalization constant. Remember that the functions $K_\nu(z)$ tend to zero exponentially in $|z|\to \infty$ and for $z>0$ they are real and positive, and therefore \eqref{phiout} is the proper solution for $\rho\geq 1$.

\item[(ii)]{Solutions in the critical and supercritical lines, corresponding to: $\mu^2=\varepsilon^2$.} 

As mentioned in the previous section, it is also necessary to analyze what happens for two concrete values of the energy $\varepsilon$ not included in (i).
These are what are usually called  critical cases  \cite{Hall,Negro22}, in which the energy takes one of the two values $\varepsilon = \pm \mu$ (the two red dashed half-lines  in Figure~\ref{ROI12}), which we will analyze below. 

$\bullet$ 
Solving the system \eqref{phis}  if $\varepsilon = \mu$, we obtain the following critical states solutions
\begin{eqnarray}
&&
\Phi^{\varepsilon = \mu}_{\ell,\rm o}(\rho,{\theta}) = a_\ell \left(
\begin{array}{c}
\rho^\ell \,  e^{i \ell \theta}
\\[1.1ex]
0
\end{array}\right),
\quad \ell\in \{-2,-3,\dots\},
\label{extreme1}
\\ [2ex]
&&
\Phi^{\varepsilon = \mu}_{\ell,\rm o}(\rho,{\theta}) = a_\ell \left(
\begin{array}{c}
\displaystyle \mu\ \rho^{-\ell} \,  e^{i \ell \theta}
\\[1.8ex]
i\, \ell \rho^{-(\ell+1)} \ e^{i (\ell+1) \theta}
\end{array}\right),
\quad \ell\in \{1,2,\dots\}.
\label{extreme2}
\end{eqnarray}
$\bullet$ 
On the other hand, when $\varepsilon = -\mu$ (sometimes these are called  supercritical states  \cite{Hall,Negro22}), the only physically acceptable solutions are
\begin{eqnarray}
&&
\Phi^{\varepsilon = -\mu}_{\ell,\rm o}(\rho,{\theta}) = a_\ell \left(
\begin{array}{c}
(\ell+1)\rho^\ell \,  e^{i \ell \theta}
\\[1.5ex]
\displaystyle -i\, \mu\ \rho^{\ell+1} \ e^{i (\ell+1) \theta}
\end{array}\right),
\quad \ell\in \{-2,-3,\dots\},
\label{extreme3}
\\ [2ex]
&&
\Phi^{\varepsilon = -\mu}_{\ell,\rm o}(\rho,{\theta}) = a_\ell \left(
\begin{array}{c}
0
\\[1.5ex]
i\ \rho^{-(\ell+1)} \ e^{i (\ell+1) \theta}
\end{array}\right),
\quad \ell\in \{1,2,\dots\}.
\label{extreme4}
\end{eqnarray}
Note that since the normalization condition \eqref{normalization} must be satisfied, for $\ell=-1,0,1$ there are no physically acceptable critical or supercritical solutions.
\vspace{2mm}
\end{itemize}
{Later, when we consider an example, we will see how a critical point for which $\varepsilon = \mu>0$ can be understood as the capture by the quantum dot of a new bound state arising from the continuous part of the spectrum just for that value of energy, while a supercritical point $\varepsilon = -\mu<0$ can be interpreted as the disappearance of a bound state in the negative sea \cite{Negro22,Peeters18}.}
\vspace{2mm}


\subsection{Solutions inside the quantum dot ($0\leq \rho\leq 1$)}\label{insideqdot}

As we know, in the model we are dealing with \eqref{potentialwell} inside the quantum dot there is a well of potential given by a depth $v$. Therefore, in this case the solutions are determined by the value of the mass-energy term of the equation (\ref{phi_1}) inside the dot: $\mu^2_{\rm i}- (\varepsilon -v_{\rm i})^2 \equiv (\lambda \mu)^2- (\varepsilon -v)^2$.
Thus, we have three different situations, depending on the sign of this term, plus a case that we will call {\it ``extreme''}, which is the one that occurs when said term is equal to zero.
Below we will analyze these three situations separately in detail:

\begin{itemize}
\item[(i)]{Solutions in the parameter region $RI_1$, characterized by  $ (\lambda \mu)^2 < (\varepsilon -v)^2$.}

This restriction that mass and energy must meet is equivalent to the inequality:
\begin{equation}\label{I1}
 \ 0<|\lambda|\mu< |\varepsilon-v|.
\end{equation}
Under the condition (\ref{I1}) in $RI_1$ (see Figure~\ref{ROI12}), the two linearly independent solutions of (\ref{phi_1}) inside the quantum dot are given in terms of the Bessel functions of the first and second kinds, $J_{\ell}$ and $Y_{\ell}$, but only the functions $J_{\ell}$ do not diverge at the origin and thus the acceptable physical solutions of \eqref{phi_1}--\eqref{phi_2} and \eqref{phi} are given by
\begin{equation}\label{phiin}
\Phi_{\ell,\rm i}(\rho,{\theta}) = b_\ell \left(
\begin{array}{c}
\sqrt{(\varepsilon-v)+(\lambda\mu)}J_\ell \left(  \rho\sqrt{ (\varepsilon-v)^2-(\lambda\mu)^2}\,\right) \  e^{i \ell \theta}
\\[1.5ex]
\displaystyle  i \, \sqrt{(\varepsilon-v)-(\lambda\mu)} \, 
J_{\ell+1} \left( \rho\sqrt{(\varepsilon-v)^2-(\lambda\mu)^2}\, \right) \ e^{i (\ell+1) \theta}
\end{array}\right),
\quad \ell\in \mathbb{Z},
\end{equation}
being $b_\ell$ normalization constants.

\item[(ii)]{Solutions in the parameter region $RI_2$, characterized by $ (\lambda \mu)^2> (\varepsilon -v)^2$.}

In this case, the values of mass and energy that allow the existence of bound states must satisfy the inequalities:
\begin{equation}\label{I2}
-|\lambda|\mu< \varepsilon-v <|\lambda|\mu.
\end{equation}
Two linearly independent solutions of \eqref{phi_1} for the parameters of region $RI_2$  (see Figure~\ref{ROI12}) are the modified Bessel functions  of the first and second kinds, $I_\ell(z)$ and $K_\ell(z)$ \cite{abramowitz}, but only the first one is acceptable, since as $z\to 0$ only $I_\ell(z)$ remains bounded. Thus, the physically acceptable solutions in $0\leq \rho\leq 1$ are
\begin{equation}\label{phiinm}
\Phi_{\ell,\rm i}(\rho,{\theta}) = b_\ell \left(
\begin{array}{c}
 \sqrt{(\varepsilon-v)+(\lambda\mu)}I_\ell \left(  \rho\sqrt{ (\lambda\mu)^2-(\varepsilon-v)^2}\,\right) \  e^{i \ell \theta}
\\[1.5ex]
\displaystyle  - i \, \sqrt{(\varepsilon-v)-(\lambda\mu)} \, 
I_{\ell+1} \left( \rho\sqrt{ (\lambda\mu)^2-(\varepsilon-v)^2}\, \right) \ e^{i (\ell+1) \theta}
\end{array}\right),
\ \ell\in \mathbb{Z}.
\end{equation}
Note the presence of the minus sign in front of the second component of the spinor in (\ref{phiinm}), as opposed to the positive signs in the rest of the solutions; this detail will have
important consequences later.

\item[(iii)]{Solutions in the parameter extreme lines  $(\lambda\mu)^2=(\varepsilon-v)^2$.} 

Here, the possible values of $\varepsilon$ to have bound states are restricted in the plane $\mu$--$\varepsilon$  to the two   half-lines 
\begin{equation}\label{linesOI}
\varepsilon =\pm |\lambda|\mu+ v=\pm \lambda \mu+v
\end{equation}
in the half-plane $\mu >0$. Hereinafter, the possible solutions $\varepsilon$ will be called {\it extreme} points or values.

 From the system of differential equations \eqref{phis} physically acceptable solutions can be obtained for these cases,
which obviously can be of two different types, depending on the $\pm$ sign in \eqref{linesOI}:

\begin{itemize}
\item[$\star$]
If $\varepsilon = \lambda\mu+ v$, then
\begin{equation}\label{c1}
\Phi^+_{\ell\geq 0,\rm i}(\rho,{\theta}) = b_\ell \left(
\begin{array}{c}
 \rho^{\ell} \  e^{i \ell \theta}
\\[1.2ex]
0
\end{array}\right)\!,
\qquad
\Phi^+_{\ell<0,\rm i}(\rho,{\theta}) = b_\ell \left(
\begin{array}{c}
\displaystyle \lambda\mu \rho^{-\ell} \  e^{i \ell \theta}
\\[1.8ex]
i \ell\rho^{-(\ell+1) } \  e^{i (\ell+1) \theta}
\end{array}\right)\!.
\end{equation}

\item[$\star$]
If $\varepsilon = -\lambda\mu+ v$, then
\begin{equation}\label{c2}
\hskip-0.6cm
\Phi^-_{\ell>0,\rm i}(\rho,{\theta}) = b_\ell \left(
\begin{array}{c}
(\ell+1) \rho^{\ell} \  e^{i \ell \theta}
\\[1.2ex]
\displaystyle- i\lambda\mu\rho^{\ell+1} \  e^{i (\ell+1) \theta}
\end{array}\right)\!\!,\
\ 
\Phi^-_{\ell\leq -1,\rm i}(\rho,{\theta}) = b_\ell \left(
\begin{array}{c}
0
\\[1.5ex]
i\rho^{-(\ell+1)} \    e^{i (\ell+1) \theta}
\end{array}\right)\!\!.
\end{equation}
\end{itemize}

\end{itemize}

\begin{figure}[htb]
\begin{center}
  \includegraphics[width=0.47\textwidth]{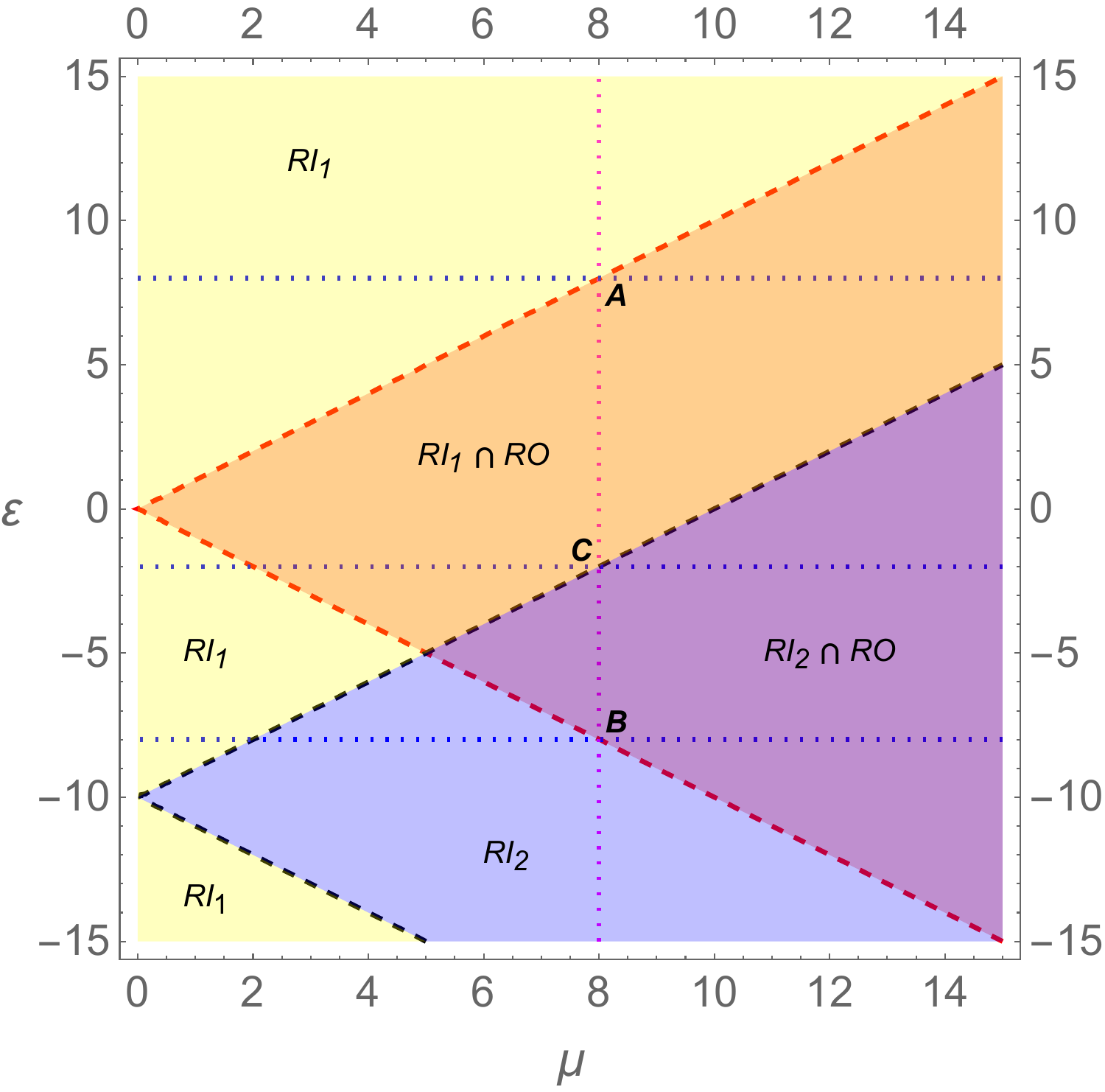}\qquad
    \includegraphics[width=0.47\textwidth]{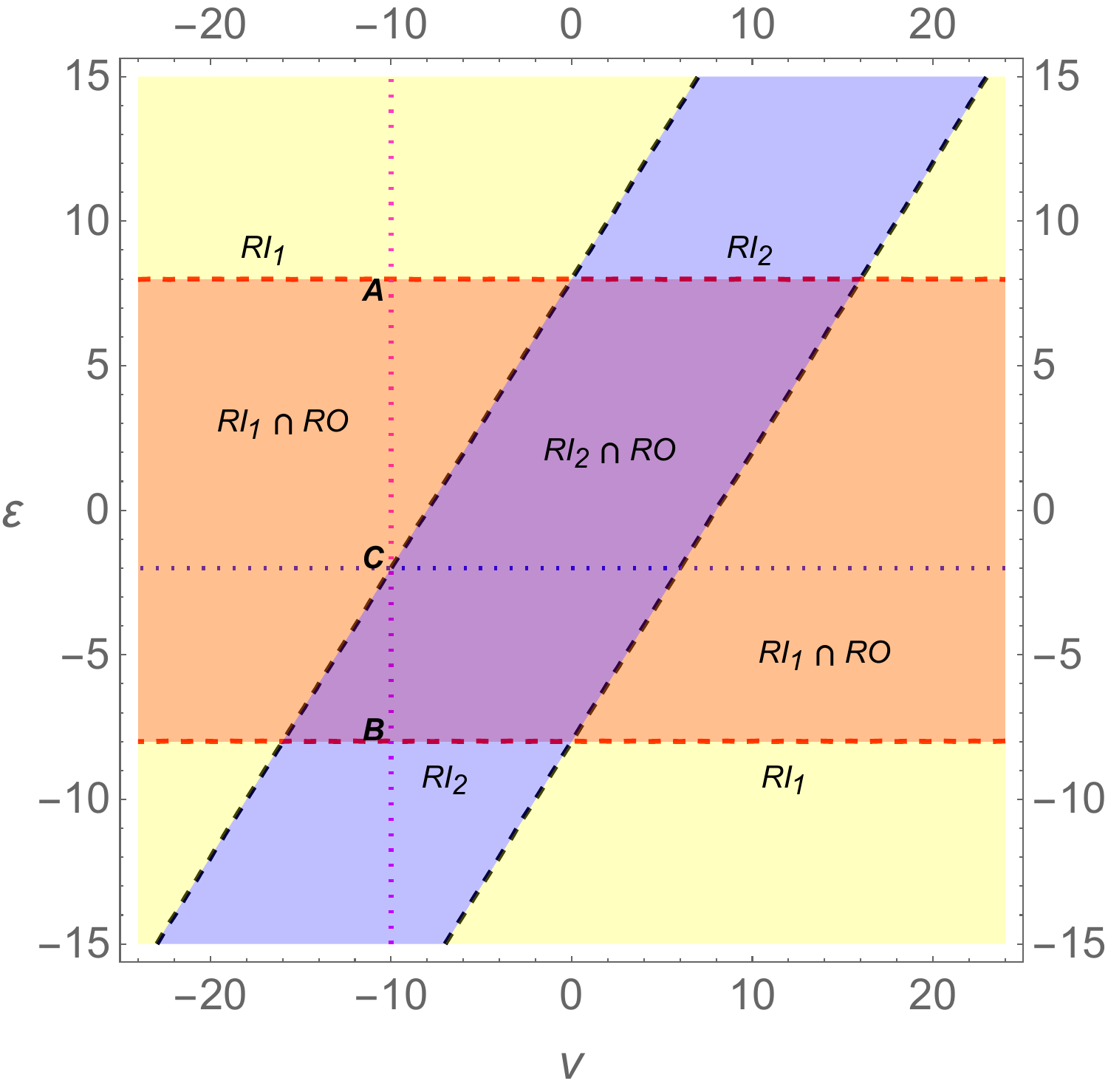}
     \caption{\small 
Graphs of the regions $RO$, $RI_1$ and $RI_2$,
where there are different forms of the acceptable solutions for the quantum dot we are studying \eqref{potentialwell}. In the drawing on the left, the value $v=-10$ has been chosen for the potential inside the point and the $\varepsilon$-$\mu$ plane has been represented, while in the drawing on the right we have assumed that $\mu=8$ and what has been represented is an $\varepsilon$-$v$ plane. In both graphs, the two dashed black rays are the {\it extreme lines} and the two dashed red rays are the {\it critical and supercritical lines} $\varepsilon=\pm\mu$. For simplicity, in both drawings we have used $\lambda=1$ and  highlighted the ``critical'' and ``supercritical'' points $A$ and $B$, and the ``extreme'' point~$C$.}
  \label{fsectors}
  \end{center}
\end{figure}

{
To illustrate what we have just explained so far, Figure~\ref{fsectors} schematically represents the regions $RO$, $RI_1$ and $RI_2$ for some typical values of the physical parameters we are handling, highlighting two specific values: $\mu=8$ (dotted magenta vertical line on the left) and  $v=-10$ (dashed magenta vertical line on the right). 
In the two graphs of Figure~\ref{fsectors}, the two black dashed lines correspond to the {\it extreme lines} defined above, and the two red dashed lines are the {\it critical} ($\varepsilon=+\mu$) and {\it supercritical} ($\varepsilon=-\mu$) lines. 
In the same drawings, the points called $A$, $B$ and $C$ have been highlighted, which are the intersection of either $\mu=8$ (left figure) or $v=-10$ (right figure) with the critical lines ($A$ and $B$) and with one of the extreme lines ($C$)  and they will play an important role, as we will see in the following sections.}

\subsection{Matching conditions at the boundary of the dot}
While we have obtained the expressions of the physically acceptable solutions for the inside and outside regions, we need to impose the continuity conditions at the quantum dot limit, at point $\rho=1$.
\newline
As we have seen, the solutions outside the quantum dot can be of two different types given by (\eqref{phiout} and \eqref{extreme1}-\eqref{extreme4}). Similarly, there are three types of  solutions in the inside region given by (\eqref{phiin}, \eqref{phiinm} and \eqref{c1}--\eqref{c2}). That way we are going to obtain a variety of secular equations, each asociated to the parameter  ($\varepsilon, v, \mu, \lambda$) regions illustrated in \ref{fsectors}:

\begin{itemize}
\item[(i)]  

Solutions within the region $RI_1 \cap RO$ in Figure \ref{fsectors}:  Given by the matching of \eqref{phiout} and \eqref{phiin} we obtain the following transcendental equation:
\begin{equation}\label{matching}
\frac{\sqrt{\mu^2- \varepsilon^2}}{\mu+ \varepsilon} \, 
\frac{K_{\ell+1} (  \sqrt{\mu^2- \varepsilon^2}\,)}{K_\ell(  \sqrt{\mu^2- \varepsilon^2}\,)}   
=
  \frac{\sqrt{(\varepsilon-v)^2-(\lambda\mu)^2 }}{(\varepsilon-v)+(\lambda\mu) } \, 
\frac{J_{\ell+1} \left( \sqrt{ (\varepsilon-v)^2-(\lambda\mu)^2}\, \right) }{J_\ell \left(   \sqrt{ (\varepsilon-v)^2- (\lambda\mu)^2}\,\right)}
\end{equation}
This secular equation gives us the allowed energy values that lead to physically 
admissible bound states inside the region $RI_1 \cap RO$. 

\item[(ii)] 
Solutions within region $RI_2 \cap RO$ in Figure \ref{fsectors}: Given by the matching between \eqref{phiout} and \eqref{phiinm}, we obtain the following transcendental equation:
\begin{equation}\label{matching2}
\frac{\lambda\mu +\varepsilon-v}{\mu+ \varepsilon}
=  - \frac{   \sqrt{ (\lambda\mu)^2 -(\varepsilon-v)^2 } }
  {\sqrt{\mu^2- \varepsilon^2 }} \, 
\frac{K_\ell(  \sqrt{\mu^2- \varepsilon^2}\,)\ I_{\ell+1} \left(  \sqrt{ (\lambda\mu)^2 -(\varepsilon-v)^2}\right)}{K_{\ell+1} (  \sqrt{\mu^2- \varepsilon^2}\,)\ I_\ell \left(   \sqrt{ (\lambda\mu)^2 -(\varepsilon-v)^2}\,\right)}
\end{equation}
This secular equation gives us the allowed energy values that lead to physically 
admissible bound states inside the region $RI_2 \cap RO$. 
\vspace{2mm}

Taking a closer look to this equation we are able to obtain parameter restrictions associated with this problem. 
\begin{itemize}
    \item[] On the one hand, since the modified Bessel functions always take positive values, the right-hand side term of this equation will always be negative. 
    \item[] On the other hand, the sign of the left-hand side term of the equation is given by the sign of $\lambda$ due to the restrictions considered to define the region $RI_2 \cap RO$. 
\end{itemize}
This is a rather peculiar situation, since it means that the effective mass inside the quantum dot $\mu_{\rm i}=\lambda\mu$ must be taken negative. We will analyze later the consequences that derive from this result.  
\item[(iii)]  Extreme solutions, defined by the black dashed lines in Figure \ref{fsectors}:
\newline
This solutions are associated with the parameter values that satisfy $ \lambda^2\mu^2=( \varepsilon-v)^2$.
As shown in \eqref{c1} and \eqref{c2}, the solutions inside the dot associated to this parameter values have four distinct expressions, based on the sign of $\ell$ as well as the energy, where each will lead to a different matching equation.
\vspace{2mm}

Out of the four resulting matching equations, only two allow for possible bound states:

$\bullet$ Solutions at 
$\varepsilon=\lambda \mu +v$: They should match spinors \eqref{phiout}  with \eqref{c1}. Nontrivial solutions may exists only for negative angular momentum values $\ell\in\{-1,-2,\dots\}$ if the following matching equation is satisfied:
    \begin{equation}\label{extr1}
  \frac{\ell}{\lambda \mu}= \frac{\mu-(v+\lambda \mu )}{\sqrt{\mu^2-(v+\lambda \mu )^2  }}   \frac{K_{\ell+1} (  \sqrt{\mu^2-(v+\lambda \mu )^2  }\,)}{K_{\ell} (  \sqrt{\mu^2-(v+\lambda \mu )^2  }\,)}
 \end{equation}
 \vspace{4mm}
 
$\bullet$ Solutions at 
$\varepsilon=-\lambda \mu +v$: They should match spinors \eqref{phiout}  with \eqref{c2}.
 Nontrivial solutions may exists only for positive angular momentum values $\ell\in\{1,2,\dots\}$ 
 if the following matching equation is satisfied:
    \begin{equation}\label{extr2}
  \frac{- \lambda \mu }{\ell+1}= \frac{\mu-(v-\lambda \mu )}{\sqrt{\mu^2-(v-\lambda \mu )^2  }}   \frac{K_{\ell+1} (  \sqrt{\mu^2-(v-\lambda \mu )^2  }\,)}{K_{\ell} (  \sqrt{\mu^2-(v-\lambda \mu )^2  }\,)}
 \end{equation}
 \vspace{2mm}
 
 Following the same logic exposed in (ii) we can see how the right-hand side term of this equations can only take positive values which combined with the angular momentum restrictions associated with each case means that only negative $\lambda$ values may lead to physically acceptable solutions. 

\item[(iv)] Critical solutions, defined by the upper red dashed line in Figure~\ref{fsectors}: 
This solutions are associated with the parameter values that satisfy $ \mu=\varepsilon$:
Similarly to the prior case,
the solutions outside the dot have two distinct expressions based on the sign of $\ell$. 
\vspace{2mm}

$\bullet$ For negative angular momentum values  $\ell\in\{-2,-3,\dots\}$: They should match spinors \eqref{extreme1} with\eqref{phiin}. Nontrivial solutions may exist if the  the following matching equation is satisfied:
  \begin{equation}\label{critn}
       J_{\ell+1} \left( \sqrt{(\mu-v)^2-(\lambda\mu)^2}\, \right)=0
  \end{equation}
  which gives  
\begin{equation}
(1-\lambda^2) \mu -2v\mu+v^2=j_{\ell+1,n}^2
\end{equation}
where $j_{\ell+1,n}$, $n=1,2,\dots$, denotes the $n$-th zero of the Bessel functions of the first kind $J_{\ell+1}(z)$. 

$\bullet$
For positive angular momentum values  $\ell\in\{1,2,\dots\}$: They should match spinors \eqref{extreme2} with\eqref{phiin}. Nontrivial solutions may exist if the  the following matching equation is satisfied:
\begin{equation}\label{critp}
\ell\left(\lambda+1-\frac{v}{\mu} \right)=\frac{\sqrt{(\mu-v)^2-(\lambda\mu)^2}\,J_{\ell+1} \left( \sqrt{(\mu-v)^2-(\lambda\mu)^2}\, \right)}{J_\ell \left(  \sqrt{ (\mu-v)^2-(\lambda\mu)^2}\,\right)}
\end{equation}
As can be verified numerically, given a set of $\ell,\lambda,\mu$ values we can always obtain at least one potential value $v$ that satisfy the prior equation. Due to the trigonometric behavior of the left hand side term.  

\item[(v)] Supercritical solutions,defined by the lower red dashed line in Figure~\ref{fsectors}: This solutions are associated with the parameter values that satisfy $  \mu=-\varepsilon$.
Similarly to the prior case,
the solutions outside the dot have two distinct expressions based on the sign of $\ell$.  
They should match spinors \eqref{extreme3}--\eqref{extreme4} with \eqref{phiin}.

$\bullet$ For negative angular momentum values $\ell\in\{-2,-3,\dots\}$: They should match spinors \eqref{extreme3} with \eqref{phiin}. Nontrivial solutions may exist if the following matching equation is satisfied:
\begin{equation}\label{supercritn}
(\ell+1)\left(\lambda-1-\frac{v}{\mu}\right)=\frac{ \sqrt{(\mu+v)^2-(\lambda\mu)^2} \, 
J_{\ell} \left( \sqrt{(\mu+v)^2-(\lambda\mu)^2}\, \right)}{J_{\ell+1} \left(  \sqrt{ (\mu+v)^2-(\lambda\mu)^2}\,\right)}
 \end{equation}
Similarly to the prior case, due to the trigonometric behavior of the left hand side term of this equation, we can always obtain at least one  we can always obtain at least one potential value $v$ that satisfy the prior equation.

$\bullet$ For positive angular momentum values $\ell\in\{1,2,\dots\}$: They should match spinors \eqref{extreme4} with \eqref{phiin}. Nontrivial solutions may exist if the following matching equation is satisfied:
\begin{equation}\label{supercritp}
J_{\ell} \left( \sqrt{(\mu+v)^2-(\lambda\mu)^2}\, \right)= 0
\end{equation}

which gives
\begin{equation}
(1-\lambda^2) \mu +2v\mu+v^2=j_{\ell
,n}^2
\end{equation}
where $j_{\ell,n}$, $n=1,2,\dots$, denotes the $n$-th zero of the Bessel functions of the first kind $J_{\ell}(z)$. 
\end{itemize}



\end{document}